# Automated Methods for Detection and Classification Pneumonia based on X-Ray Images Using Deep Learning

**Khalid EL ASNAOUI[1], Youness CHAWKI[2], Ali IDRI[1, 3]**

[1] Complex system engineering and human system, Mohammed VI Polytechnic University
Benguerir, Morocco

[2] Moulay Ismail University, Faculty of Sciences and Techniques
Errachidia, Morocco

[3] Software Project Management Research Team, ENSIAS, University Mohammed V in Rabat, Morocco

khalid.elasnaoui@um6p.ma, youness.chawki@gmail.com, ali.idri@um6p.ma

**Abstract:** Recently, researchers, specialists, and companies around the world are rolling out deep learning and image processing-based systems that can fastly process hundreds of X-Ray and computed tomography (CT) images to accelerate the diagnosis of pneumonia such as SARS, COVID-19, and aid in its containment. Medical images analysis is one of the most promising research areas, it provides facilities for diagnosis and making decisions of a number of diseases such as MERS, COVID-19. In this paper, we present a comparison of recent Deep Convolutional Neural Network (DCNN) architectures for automatic binary classification of pneumonia images based fined tuned versions of (VGG16, VGG19, DenseNet201, Inception_ResNet_V2, Inception_V3, Resnet50, MobileNet_V2 and Xception). The proposed work has been tested using chest X-Ray & CT dataset which contains 5856 images (4273 pneumonia and 1583 normal). As result we can conclude that fine-tuned version of Resnet50, MobileNet_V2 and Inception_Resnet_V2 show highly satisfactory performance with rate of increase in training and validation accuracy (more than 96% of accuracy). Unlike CNN, Xception, VGG16, VGG19, Inception_V3 and DenseNet201 display low performance (more than 84% accuracy).

## I. Introduction

Throughout history, epidemics and chronic diseases have claimed the lives of many people and caused major crises that have taken a long time to overcome. To describe a disease within populations that arise over a specific period of time, two words are used epidemic and outbreak (Obrane et al., 2017; CDCP, 2012). Indeed, we can define epidemic as the occurrence of more cases of illnesses, injury or other health condition than expected in a given area or among a specific group of persons during a particular period. Mostly, the cases are pretending to have a common cause (CDCP, 2012). The outbreak is distinguished from an epidemic as more localized, or the term less likely to evoke public panic.

Past epidemics include pneumonia. The pneumonia is an infection of the lungs most often caused by a virus or bacteria. More specifically, the infection affects the pulmonary alveoli, the tiny balloon-shaped sacs at the end of the bronchioles (see figure 1). It usually affects only one of the 5 lobes of the lung (3 lobes in the right lung and 2 in the left), hence the term lobar pneumonia. When pneumonia also reaches the bronchial tubes, it is called "Bronchopneumonia". It is the most important cause of death in the world for children younger than 5 years (about 12.8% of annual deaths) (PERCH, 2019; Liu et al., 2016). It is also a leading cause of morbidity and mortality in adults worldwide and in particular in China (Tian et al., 2020; Prina et al., 2015; Welte et al., 2012).Pneumonia is the third leading cause of death in Japan with a higher mortality rate for the elderly, particularly among individuals ≥80 years old (Kondo et al., 2017). Excluding lung cancer, in Portugal, Pneumonia is the huge cause of respiratory death (Hespanhol and Barbara, 2017).

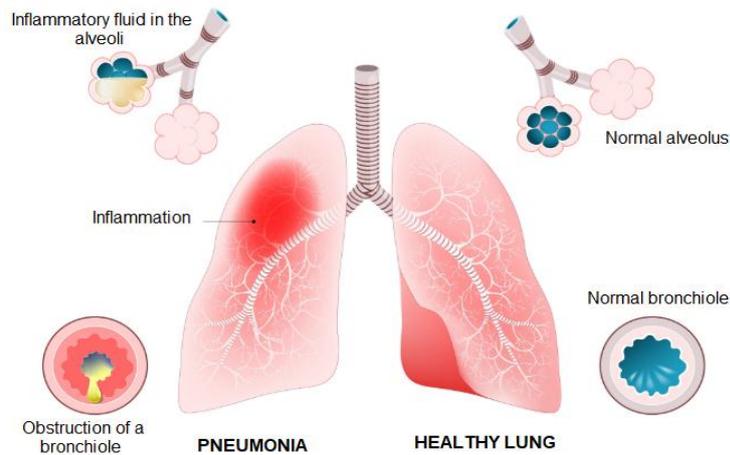

**Fig. 1**: Pneumonia Diagram

Several Coronavirus have passed over the species barrier to cause deadly pneumonia in humans since the beginning of the 21st century. To know the pathogenesis of these deadly epidemics, the specialists have to examine the structure of the viruses and the mechanism of infection. This allows them to help elucidate and provide information for the development of effective treatment and possibly vaccines (Yang et al., 2020). Here, we will provide a brief summary of the epidemiology and history of type of Coronavirus in particular: SARS, MERS and Covid-19. Table 1 shows the major pandemics that have occurred over time:

**Table1**: Major pandemics that have occurred over time

| Name | Time period | Type/Pre-human host | Death toll |
|---|---|---|---|
| Antonine Plague | 165-180 | Believe to either smallpox or measles | 5M |
| Japanese smallpox epidemic | 735-737 | Variola major virus | 1M |
| Prague of Justinian | 541-542 | Yersinia pestis bacteria/Rats, fleas | 30-50M |
| Black death | 1347-1351 | Yersinia pestis bacteria/Rats, fleas | 200M |
| New World Smallpox Outbreak | 1520-onwards | Variola major virus | 56M |
| Great Plague of London | 1665 | Yersinia pestis bacteria / Rats, fleas | 100 000 |
| Italian plague | 1629-1631 | Yersinia pestis bacteria / Rats, fleas | 1M |
| Cholera Pandemics 1-6 | 1817-1923 | V. cholerae bacteria | 1M+ |
| Third Plague | 1985 | Yersinia pestis bacteria / Rats, fleas | 12M (China and India) |
| Yellow Fever | Late 1800s | Virus / Mosquitos | 100 000-150 000 (US) |
| Russian Flu | 1889-1890 | Believed to be H2N2 (avian origin) | 1M |
| Spanish Flu | 1918-1919 | H1N1 virus / Pigs | 40-50M |
| Asian Flu | 1957-1958 | H2N2 virus | 1.1M |
| Hong Kong Flu | 1968-1970 | H3N2 virus | 1M |
| HIV/AIDS | 1981- | Virus / Chimpanzees | 25-35M |

|  | Present |  |  |
| --- | --- | --- | --- |
| Swine Flu | 2009-2010 | H1N1 virus / Pigs | 200 000 |
| SARS | 2002-2003 | Coronavirus / Bats, Civets | 774 |
| Ebola | 2014-2016 | Ebolavirus / Wild animals | 11 000 |
| MERS | 2015-Present | Coronavirus / Bats, Camels | 850 |
| Covid-19 | 2019-Present | Coronavirus – Unknown (possibly Bats or pangolins) | 34,845 (as of Mar 28, 2020) |

SARS-Cov (Severe Acute Respiratory Syndrome Coronavirus) (Drosten et al., 2003; Ksiazek et al., 2003) is an acute respiratory illness caused by a coronavirus, characterized by fever, coughing, breathing difficulty, and usually pneumonia. SARS It appeared first time in China exactly in the province of Guangdong in 2002 and spread to the world through airtravel routes. Approximately 8098 people were affected, causing in 774 deaths (Walls et al., 2020; Roussel et al., 2020; Wit et al., 2016) with a lethal rate about 10% (Wu et al., 2020). It is suggested to originate from bats (Walls et al., 2020; Hu et al., 2017). SARS symptoms are usually same to flu symptoms: fever, chills, muscle aches, headache and occasionally diarrhoea. After about one-week, other symptoms appear like fever of 38°C or higher, dry cough, breath shortness (Wu et al., 2020).

MERS-Cov (Middle East Respiratory Syndrome Coronavirus) a viral respiratory illness caused by a virus (WHO, 2018). First, it was appeared in the Middle East and exactly Saudi Arabia in 2012 (Van Boheemen et al., 2012, Zaki et al., 2012). Other cases were identified in Jordan (Hijawi, 2013), Qatar (Farooq et al., 2020) then spread to the world. MERS is a zoonotic virus that can be transmitted between animals and people. Indeed, the World Health Organization has confirmed that humans are affected because they were on contact with affected dromedary camels (WHO, 2019; Mohd et al., 2016; Azhar et al., 2014). Studies have shown that how the virus is transmitted from animals to humans is not yet understood, and also the human-to-human transmission is very limited unless there is close contact (WHO, 2018; Ki, 2015; Oboho et al., 2015). The different MERS symptoms are as follows: Fever, Cough (Dry, Productive), Shortness of breath, Diarrhea, Myalgia, Headache, Nausea, Vomiting, Abdominal pain, Chest pain, Sore throat, Haemoptysis (Farooq et al., 2020; Wang et al., 2019; Baharoon et Memish, 2019; Al-Omari et al., 2019; WHO, 2018; Aguanno et al., 2018). The world is currently experiencing a dangerous viral epidemic caused by a virus that has killed tens of thousands of people. This new virus called Covid-19 was identified in Wuhan, China in December 2019 (Guo et al., 2020; Lippi et al. 2020; Amrane et al., 2020; Cortegiani et al., 2020; Tian et al., 2020; Tang et al., 2020; Faralli et al., 2020; Wilder-Smith et al., 2020; Driggin et al. 2020; Cheng et al., 2020; Pung at al., 2020; Zhang et al., 2020; El Zowalaty and Järhult, 2020). It belongs to the Corona family of viruses, but it is more deadly and dangerous than the rest of the coronaviruses (Li et al., 2020; WHO, 2020). First cases of the disease have been related to a live animal seafood market in Wuhan, denoting to a zoonotic origin of the epidemic (Kandel et al., 2020; Sun et al., 2020; El Zowalaty and Järhult, 2020; Wilder-Smith et al., 2020; Roosa et al., 2020; CNHC, 2020). The routes of transmission, treatments, and results of Covid-19 continually receiving much research attention in the world (Guo et al., 2020). Indeed, researchers have identified three main modes of the virus transmission: close person-to-person contact, aerosol transmission and transmission by touch. (Yang et al., 2020; Lu et al., 2020; Li et al., 2020; Liu et al., 2020). What makes the Corona virus very dangerous is that it can have up to two weeks of incubation without symptoms. We can cite the symptoms of Covid-19: high fever, dry cough, tiredness, shortness of breath, aches and pains, sore throat and very few people will report diarrhoea, nausea or a runny nose (Simcock et al., 2020; Yang et al., 2020; WHO, 2020).

As the number of patients infected by this disease increases, it turns out to be increasingly hard for radiologists to finish the diagnostic process in the constrained accessible time (Zhang et al., 2011). Medical images analysis is one of the most promising research areas, it provides facilities for diagnosis and making decisions of a number of diseases such as MERS, COVID-19. Recently, many efforts and more attention are paid to imaging modalities and DL in pneumonia. Therefore, interpretation of these images requires expertise and necessitates a several algorithms in order to enhance, accelerate and make an accurate diagnosis. Following this context, DL algorithms (Bhandary et al., 2020) have obtained better performance in the detection of pneumonia, and demonstrated high accuracy compared with previous state of the art methods. Motivated by the fastest and accurate detection rate of pneumonia using deep learning (DL), our work is going to present a comparison of recent Deep Convolutional Neural Network (DCNN) architectures for automatic binary classification of X-Ray images between normal and pneumonia in order to answer to the following research question: Is there any Deep Learning techniques which distinctly outperforms other DCNN techniques?

The contributions of our paper are as followings: (1) We design fined tuned versions of (VGG16, VGG19, DenseNet201, Inception_ResNet_V2, Inception_V3, Xception, Resnet50, and MobileNet_V2), (2) To avoid overfitting in our models, we used weight decay and regularizers that are provided under keras and have the name L2. Our models have been tested on the chest X-Ray & CT dataset (Kermany et al., 2018) for binary classification.

The remainder of this paper is organized as follow. Section 2 deals with some related work. In Section 3, we describe our proposed methods. Section 4 presents some results obtained using our proposed models and interpreting the results. The conclusion is given in the last section.

## II. Related works

To make a very good diagnosis and to detect the source of the problems of the diseases and in an oppurtum time remains a major challenge for the doctors in order to reduce the sufferings of the patients. Indeed, the use of image processing and deep learning algorithms in the analysis and processing of biomedical images has given very satisfactory results. In this section, a brief review of some important contributions from the existing literature is presented.

Pneumonia remains one of the diseases that is increasingly becoming research hotspots in recent years. Indeed, Toğaçar et al. (Toğaçar et al., 2019) used the lung X-ray images. They employed the convolutional neural network as feature extractor and used some existing convolutional neural network models like AlexNet, VGG-16 and VGG-19 to realize a specific task. The number of deep features was reduced using of deep features was reduced for each deep model. Then a step of classification was done using decision tree, KNN, linear discriminant analysis, linear regression, and SVM. The obtained results showed that the deep features provided robust and consistent features for pneumonia detection.

In (Liang and Zeng, 2020), Liang and Zeng have proposed a new deep learning framework to classify child pneumonia image bay combining residual thought and dilated convolution. To overcome the over-fitting and the degradation problems of the depth model and the problem of loss of feature space information caused by the increment in depth of the model the proposed method used the residual structure and dilated convolution respectively.

The work presented in (Jaiswal et al., 2019) proposes a deep learning-based method to identify and to localize the pneumonia in Chest X-Rays images. The identification model is based on Mask-RCNN that can incorporate global and local features for pixel-wise segmentation. Good performances evaluated on chest radiograph dataset showed the effectiveness and robustness of the model.

The investigation of post-stroke pneumonia prediction models using advanced machine learning algorithms, specifically deep learning approaches has been presented in (Ge et al., 2019). Indeed, authors have used the classical methods (logistic regression, support vector machines, extreme gradient boosting) and have implemented methods based on multiple layer perceptron neural networks and recurrent neural network to make use of the temporal sequence information in electronic health record systems. The obtained results showed that the deep learning-based predictive model achieves the optimal performance compared to many classic machine learning methods.

In (Sirazitdinov et al., 2019), authors have proposed an automated detection and localization method of pneumonia on chest x-ray images using machine learning solutions. They presented two convolutional neural networks (RetinaNet and Mask R-CNN). The proposed method was validated on a dataset of 26,684 images from Kaggle Pneumonia Detection Challenge and the obtained results were satisfactory.

Bhandary et al. (Bhandary et al., 2020) have proposed a Deep Learning framework for examining lung pneumonia and cancer. Indeed, they proposed two different deep learning techniques: the first one was Modified AlexNet. It was intended to classify chest X-Ray images into normal and pneumonia class using Support Vector Machine and its performance were validated with other pre-trained deep learning (AlexNet, VGG16, VGG19 and ResNet50). Whereas the second one implemented a fusion of handcrafted and learned features in the MAN for improving classification accuracy during lung cancer assessment.

The work presented in (Behzadi-khormouji et al., 2020) presents a method that can detect consolidations in chest x-ray radiographs using Deep Learning especially Convolutional Neural Networks assisting radiologist for better diagnosis. Authors have used a Deep Convolutional Neural Network pre-trained with ImageNet data to improve the accuracy of the models. Then, to enhance the generalization of the models, they proposed three-step pre-processing approach.

When we use DL and image processing, especially in medical imaging, a natural question arises: how we use this DL, and which is the best architecture of DL will be used? Following the context of X-Ray images classification, we present a comparison of recent Deep Convolutional Neural Network (DCNN) architectures for automatic binary classification based fined tuned versions of (CNN, VGG16, VGG19, DenseNet201, Inception_ResNet_V2, Inception_V3, Xception, Resnet50, and MobileNet_V2).

## III. Proposed contributions

Deep Learning (DL) methods have recently demonstrated huge potential with state-of-the-art performance on image processing and computer vision (Alom et al., 2018). These techniques have been applied in various modalities of medical imaging with high performance in segmentation, detection, and classification (Litjens et al., 2017). Some DL methods incorporate skin cancer detection, breast cancer detection, and classification, lung cancer detection, etc. Despite the fact that these methods have shown huge achievement in terms of success in medical imaging, they require a large amount of data, which is as yet not available in this field of applications. Following the context of no availability of medical imaging dataset, our work is going to fine-tuned (Transfer learning (Bhandary et al. 2020))) the top layer of the following DL architectures (CNN, VGG16, VGG19, DenseNet201, Inception_ResNet_V2, Inception_V3, Xception, Resnet50, and MobileNet_V2) and compare between them their performances.

### III.1. Proposed baseline CNN architecture

Generally, a CNN model consists of five layers which are: input layer, convolutional layers, pooling layers, full-connection layers, and output layer (Figure 2). Moreover, it is known that CNN model is can be trained end-to-end in order to allow the feature extraction and selection, finally, classification or prediction. How the network understands an image or process the image is difficult, but it is known that features obtained in different layers of a network works better than human built features (Ouhda et al., 2019).

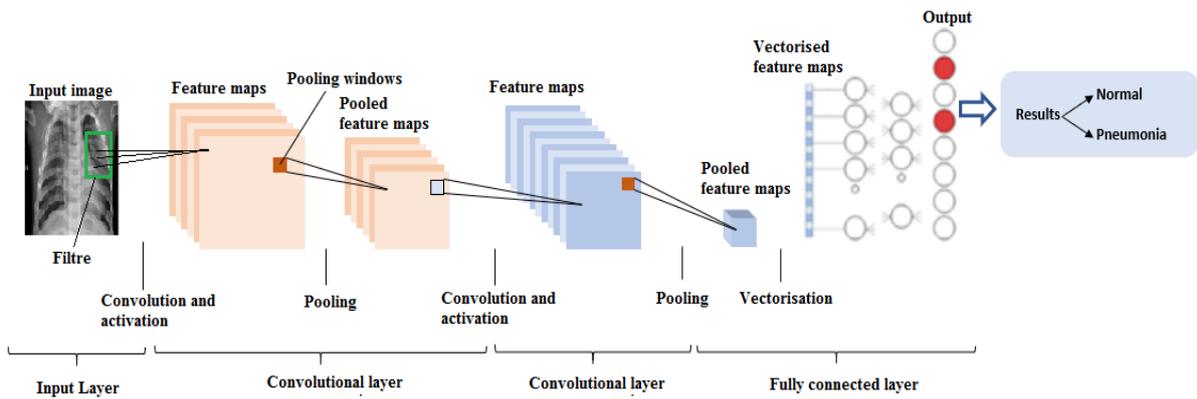

**Fig. 2**. Main architecture of our baseline CNN

The proposed baseline CNN for our experiment has the following architecture:

- Input layer: In our experiment the inputs are X-Ray images. The parameters are defining the image dimension (244x244).
- Convolutional layers: a convolution is a linear operation that consists on the multiplication of a set of weights with the input. It's designed for two-dimensional input; the multiplication is performed between a two-dimensional array of weights (filters) and an array of input data. In the proposed architecture we have 3 layers with a filter of size 3x3 and zero padding.
- Pooling Layers: Pooling layers is a technique to down sample feature maps by summarizing the presence of features in patches of the feature map. There is two type of pooling methods that are average pooling and max pooling. In the proposed architecture we used Max pooling in order to calculate the maximum value in each patch for every feature map. The max pooling is set to 2x2 with a stride of 2.
- ReLU layers: we have used 4 ReLU layer for each convolutional layer
- Inner-product layers or fully connected layers: Treats the input data as a simple vector and produce an output as a single vector. We have one inner-product layer in this model. The last one, a fully connected output layer with sigmoid activation.

### III.2. Deep Learning architectures

Deep learning architectures is highly used for the diagnosis of pneumonia since 2016 (Bhandary et al. 2020; Kermany et al. 2018), the most investigated DL techniques are VGG16, VGG19, Inception_V3, DenseNet201, Xception, Resnet50, Inception ResnetV_2, and MobileNet_V2. We have chosen these 8 techniques due to the high result and accuracies they offer.

**VGG16 and VGG19**

Proposed on 2014 by Simonyan and Zisserman VGG (Visual Geometry Group) is a convolution neural net (CNN) architecture and used to win ILSVR(ImageNet) competition in 2014 (K. Simonyan and A. Zisserman, 2014). The major characteristic of this architecture is instead of having a large number of hyperparameters, they concentrated on simple 3×3 size kernels in convolutional layers and 2×2 size in max pooling layers. In the end, it has 2 FC (Fully Connected layers) trailed by a softmax for output. The most familiar VGG models are VGG16 and VGG19 which include 16 and 19 layers, respectively. The difference between VGG-16 and VGG-19 is that VGG-19 has one more layer in each of the three convolutional blocks (Zhang et al., 2019.)

**Inception_V3**

Inception models are a type on Convolutional Neural Networks developed by Szegedy on 2014 (Szegedy et al., 2014). The inception models differ from the ordinary CNN in the structure where the inception models are inception blocks, that means lapping the same input tensor with multiple filters and concatenating their results. Inception_V3 is new version of inception model presented for the first time on 2015 (Szegedy et al., 2015). It is an improved version of inception_V1 and inception_V2 with more parameters. Indeed, it has block of parallel convolutional layers with 3 different sizes of filters (1x1, 3x3, 5x5). Additionally, 3×3 max pooling is also performed. The outputs are concatenated and sent to the next inception module.

**Resnet 50**

Resnet50 is a deep residual network developed by (He at al., 2016) and is a subclass of convolutional neural networks used for image classification. It is the winner of ILSVRC 2015. The principal innovation is the introducing of the new architecture network-in-network using residual layers. The Renset50 consists of five steps each with a convolution and Identity block and each convolution block has 3 convolution layers and each identity block also has 3 convolution layers. Resnet50 has 50 residual networks and accepts images of $224 \times 224$ pixels.

**Inception_Resnet_V2**

Inception-ResNet-v2 is a convolutional neural network that is trained on more than a million images from the ImageNet database (Szegedy et al., 2016). It is a hybrid technique combining the inception structure and the residual connection. The model accepts images of 299×299 image, and its output is a list of estimated class probabilities. The advantages of Inception ResnetV2 are converting inception modules to Residual Inception blocks, adding more Inception modules and adding a new type of Inception module (Inception-A) after the Stem module.

**Densenet201**

DenseNet201 (Dense Convolutional Network) is a convolutional neural network that is 201 layers deep and accepts an image input size of $224 \times 224$ (Huang et al., 2018). DenseNet201 is an improvement of ResNet that includes dense connections among layers. It connects each layer to every other layer in a feed-forward fashion. Unlike traditional convolutional networks with L layers that have L connections, DensNet201 has L(L+1)/ 2 direct connections. Indeed, compared to traditional networks, DenseNet has can improve the performance by increasing the computation requirement, reducing the number of parameters, encouraging feature reuse and reinforcing feature propagation.

**MobileNet_V2**

MobileNet_V2 (Sandler et al., 2018) is a convolutional neural network and an improved version of MobileNetV1 and is made of only 54 layers and has an image input size of $224 \times 224$. Its main characteristic is instead of performing a 2D convolution with a single kernel, instead of performing a 2D convolution with a single kernel uses depthwise separable convolutions that consists in applying two 1D convolutions with two kernels. That means, less memory and parameters required for training and small and efficient model. We can distinguish two types of blocks: first one is residual block with stride of 1, second one is block with stride of 2 for downsizing. For each block, there are layers: the first layer is 1×1 convolution with ReLU6, the second layer is the depthwise convolution and the third layer is another 1×1 convolution but without any non-linearity.

**Xception**

Xception, presented by Chollet (Chollet, 2017), is a convolutional neural network that is 71 layers deep. It is an improved version of Inception architecture and involves depthwise deparable convolutions. More precisely, Xception replaces the standard Inception modules with depthwise separable convolutions. It showed good results compared to VGG16, Resnet and Inception in classical classification problems. Xception has an image input size of 299x299.

## IV. Experimental results and analysis

We implemented our contributions for automatic binary classification on a new publicly available image dataset (chest X-Ray & CT dataset). As it can be observed in Figure 3 which shows the diagram of the main proposed contributions, the entire contributions are mainly divided into three steps: data acquisition, data pre-processing and classification. The following sections give out in detail the steps of our contribution.

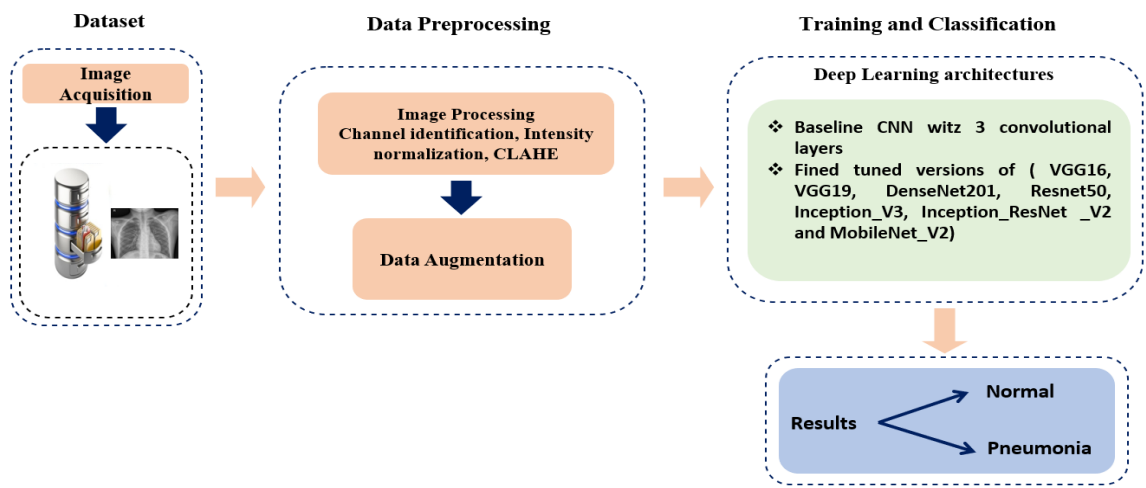

**Fig. 3**: Block diagram of the proposed contributions

### IV.1. Dataset

This present work introduces a publicly available image dataset which contains X-Ray and computed tomography (CT) images. This dataset, named chest X-Ray & CT dataset available in this link, is composed of 5856 images (jpeg format) and has two categories (4273 pneumonia and 1583 normal). As can be seen from Figure 4 that illustrates an examples of chest X-Rays in patients with pneumonia, the normal chest X-Ray (Figure 4. (a)) shows clear lungs with no zones of abnormal opacification. Moreover, Figure 4. (b) regularly shows a focal lobar consolidation (white arrows), while figure 4. (c) shows with a more diffuse "interstitial" pattern in two lungs (Kermany et al., 2018).

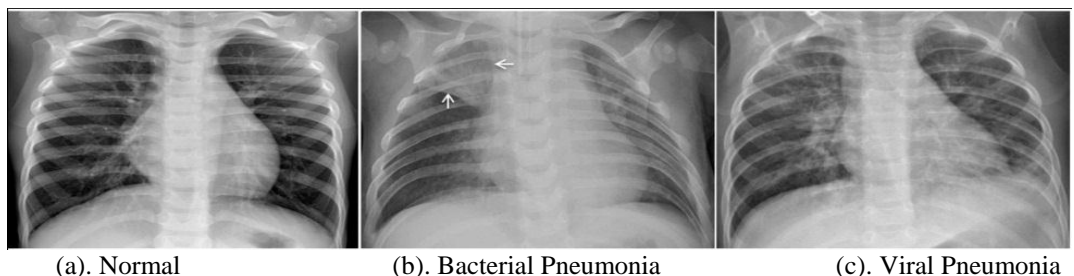

(a). Normal          (b). Bacterial Pneumonia          (c). Viral Pneumonia
**Fig. 4**: Examples of Chest X-Rays in patients with pneumonia

### IV.2. Data pre-processing and data splitting

The next stage is to pre-process input images using different pre-processing techniques. The motivation behind image pre-processing is to improve the quality of visual information of each input image (eliminate or decrease noise present in the original input image, improve image quality through increased contrast, delete the low or high frequencies, etc). In this study, we used intensity normalization and Contrast Limited Adaptive Histogram Equalization (CLAHE).

Intensity normalization is an interesting pre-processing step in image processing applications (Kassani et al., 2019). In our models, we normalized input images (Figure 5. (a)) to the standard normal distribution using min-max normalization (Equation1). Furthermore, before feeding input images into the proposed models, CLAHE is a necessary step to improve the contrast in images (Kharel et al., 2017; Makandar and Halalli, 2015). Figure 5 illustrates an example of using these techniques.

$$X_{norm} = \frac{x - x_{min}}{x_{max} - x_{min}} \qquad (1)$$

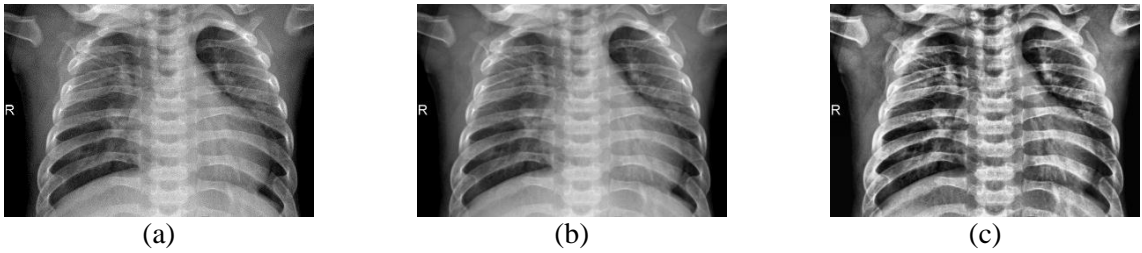

(a)        (b)        (c)

**Fig. 5**. Normalized image:
(a). Original image, (b). Normalized image, (c). CLAHE

For data splitting, we used in this experiment 60% of the images for training and 40% of the images for validation. We ensure that the images chosen for validation are not used during training in order to perform successfully the binary classification task. Moreover, we observed that our dataset is imbalanced, indeed 75% of the images represents the pneumonia class. In order to overcome this issue, we resampled our dataset by using data augmentation. In fact, we generated from each single input image 2 new images with different augmentation techniques. Therefore, the total number of images in normal class was increased by 2 times.

### IV.3. Data augmentation

Data augmentation is used for the training process after dataset pre-processing and splitting and has the goal to avoid the risk of over fitting. Moreover, the strategies we used include geometric transforms such as rescaling, rotations, shifts, shears, zooms and flips (Table 2).

**Table 2**: Data augmentation used

| Argument | Parameter value | Description |
|---|---|---|
| Rescale | 1 / 255.0 | Scale images from integers 0-255 to floats 0-1 |
| Rotation range | 90 | Degree range of the random rotations |
| Vertical shift range | 0.2 | The parameter value of horizontal and vertical shifts (20%) is a fraction of the given dimension |
| Horizontal shift range | 0.2 | |
| Shear range | 0.2 | Controls the angle in counterclockwise direction as radians in which our image will allowed to be sheared |
| Zoom range | 0.2 | Allows the image to be "zoomed out" or "zoomed in" |

| Horizontal flip | True | Controls when a given input is allowed to be flipped horizontally during the training process |
| Fill mode | Nearest | This is the default option where the closest pixel value is chosen and repeated for all the empty values |

### IV.4. Training and classification dataset

After data pre-processing, splitting and data augmentation techniques used, our training dataset size is increased and ready to be passed to the feature extraction step with the proposed models in order to extract the appropriate and pertinent features. The extracted features from each proposed model are flattened together to create the vectorized feature maps. The generated feature vector is passed to a multilayer perceptron to classify each image into corresponding classes. Finally, the performance of the proposed method is evaluated on test images using the trained model. We repeat each experiment three times and report their average results.

### IV.5. Experimental setup

Toward an automatic binary classification based on a publicly available image dataset (Chest X-Ray dataset), our experimentations were carried out based on following experimental parameters: All the images of the dataset were resized to 224x224 pixels except those of Inception_V3 model that were resized to 299x299. To train our models, we set the batch size to 32 with the number of epochs set to 300. The training and validation samples are respectively set to 159 and 109. Adam with $\beta 1=0.9$, $\beta 2=0.999$ is used for optimization, and learning rate set to 0. 00001 and decreased it to 0.000001. Moreover, we used weight decay to reduce overfitting of our models. The regularizers are provided under keras and have the name L2. The implementation of the proposed models is done using computer with Processor: Intel (R) core (TM) i7- 7700 CPU @ 3.60 GHZ and 8 Go in RAM running on a Microsoft Windows 10 Professional (64-bit). For simulation, python 3 is used and Keras/tensorflow as deep learning backend. Our training and validation steps are running using NVIDIA Tesla P40 with 24 Go RAM.

### IV.6. Evaluation criteria

After extracting the appropriate feature, the last step is to classify the attained data and assign it to a specific class (Blum et al., 2001). Among the different classification performance properties, and since our data is balanced our study uses the following benchmark metrics, including accuracy, sensitivity, specificity, precision and F1 score (Blum et al., 2001; Bhandary et al. 2020). These popular parameters are defined as follows:

$$Accuracy = \frac{TP+TN}{TN+TP+FP+FN}$$

$$Sensitivity = \frac{TP}{TP+FN}$$

$$Specificity = \frac{TN}{TN+FP} \quad (2)$$

$$Precision = \frac{TP}{TP+FP}$$

$$F1 = 2 \times \frac{Recall \times precision}{Recall + precision}$$

Where: TP: True Positive. FP: False Positive. TN: True Negative, and FN: False Negative.

### IV.7. Results and discussion

In this section we present the result for the binary classification for the chest X-Ray & CT images with the following architectures (Baseline CNN, Fine tuning the top layers of VGG16, VGG19, Inception_V3, Xception, Resnet50, Inception_Resnet_V2, DenseNet201, and MobileNet_V2). In addition, to check the performance and robustness of each proposed model, several experiments are conducted on chest X-Ray dataset. The results are presented separately (Fig.6 - 15) using training and validation curves of accuracy and loss. Other experiments are tabulated below (Table.3).

### IV.7.1. Classification results of the proposed architectures

This subsection presents and discusses of the classification results of chest X-Ray & CT images. Before to discuss these results, let us define the two most parameters used in the state of the art of deep learning and computer vision: Train curve is calculated from the training dataset that provides an idea of how well the model is learning, while the validation curve or test curve is calculated from a hold-out validation dataset that provides an idea of how well the model is generalizing. Moreover, the training and validation loss is defined as a summation of the errors made for each example in validation or training sets. In contrast to accuracy, loss is not a percentage. To summarize, a model that generalizes well is a model that is neither overfit nor underfit, is the best model. Furthermore, the confusion matrix shows the detailed representation of what happens with images after classification (Bhandary et al., 2020).

- **Baseline CNN**

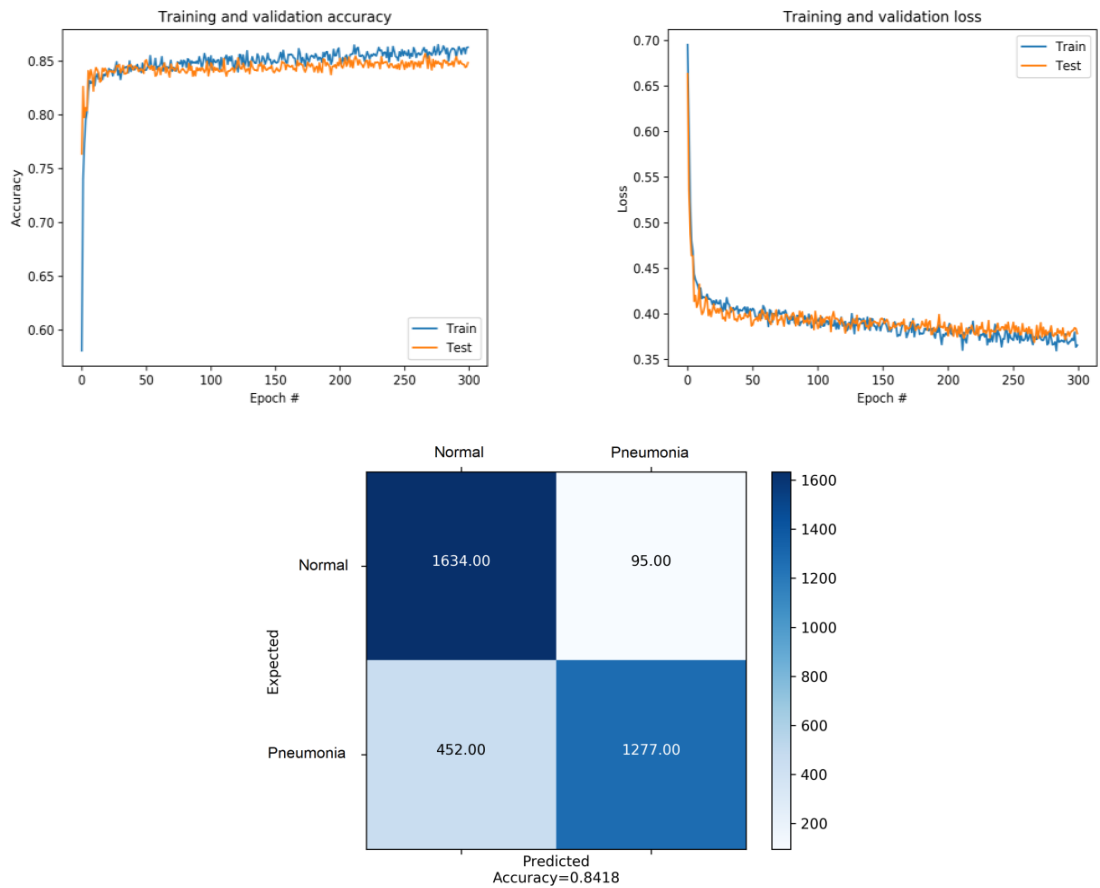

**Fig. 6**. Accuracy and loss curve and confusion matrix of Baseline CNN

According to the figure above, it is observed that the accuracy curve of train data is rapidly increasing from epoch 0 to epoch 6 where the accuracy is equal to 83.17% then it starts to be stable until epoch 300 where the accuracy is equal to 86.28% Same for the accuracy curve of test data with an accuracy of 84.84 for epoch 300.

For the loss curve of train data is rapidly decreasing from epoch 0 to epoch 6 where the loss is 43.77% then it starts to be stable until the end of training (epoch 300) where the loss is equal to 36.56%. Same for loss curve of data test with a loss of 37.85% for epoch 300.

From the confusion matrix, it is observed that the first images class (Normal), the model was able to identify 1634 images correctly in the normal class, but 95 were labelled as Pneumonia. Likewise, for the second images class (Pneumonia), the model was able to identify 1277 images correctly, but 252 images were labelled as Normal.

- **VGG16**

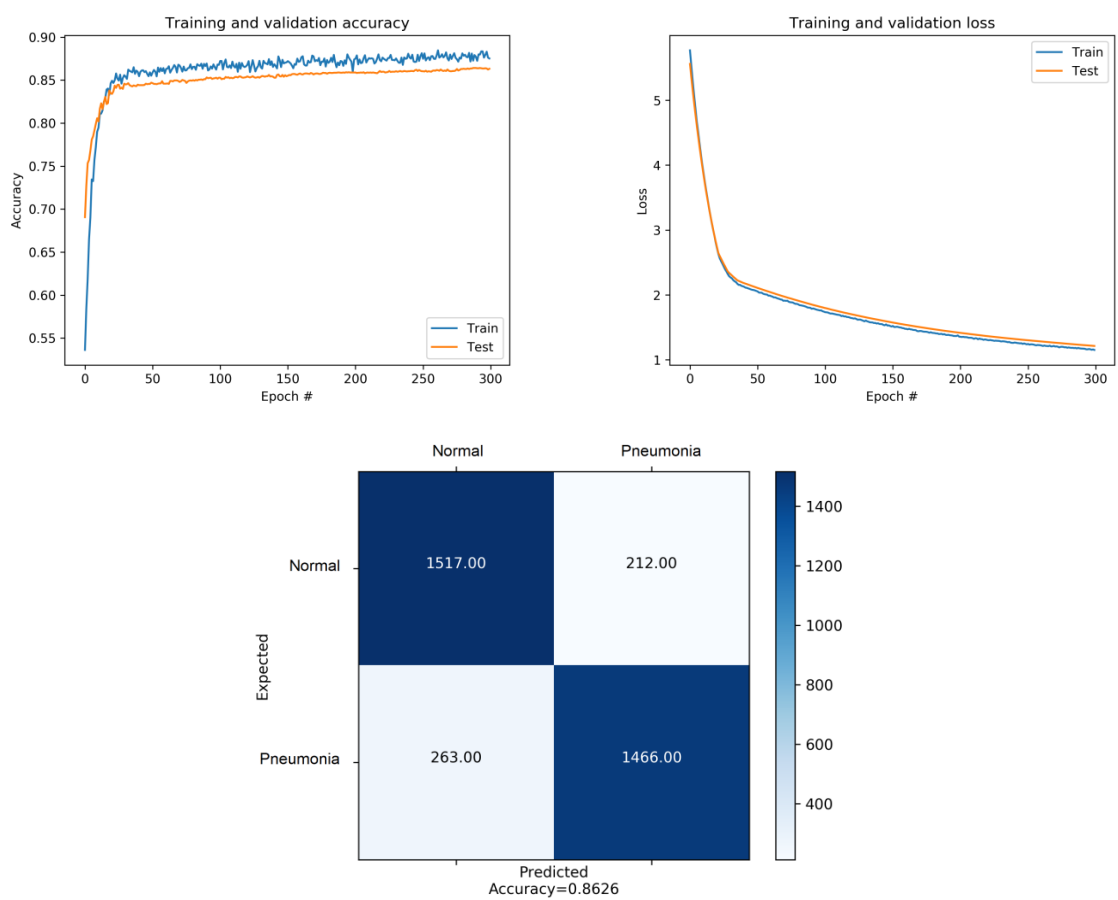

**Fig. 7**. Accuracy and loss curve and confusion matrix of VGG16

The following figure presents accuracy and loss curve and confusion matrix of VGG16. Indeed, from the epoch 0 to epoch 11, the accuracy curve of train data is quickly increasing where is equal to 81.05% then it converges to a value of 87.51%. Same for the accuracy curve of test data with an accuracy of 86.32% for epoch 300.

A rapid decreasing of loss curve can be observed for train data from epoch 0 to epoch 25 where the loss is equal to 1.43% then a kind of stability can be observed up to the value 1. 15%. The same goes for the loss curve of test data where the loss is equal to 2.21 for epoch 300.

From the confusion matrix, we can observe for images class (Normal) the model was able to predict 1517 images correctly in the normal class, but 212 were labelled as Pneumonia. For the images class

(Pneumonia), the model was able to identify 1466 images correctly, but 263 images were labelled as Normal.

- **VGG19**

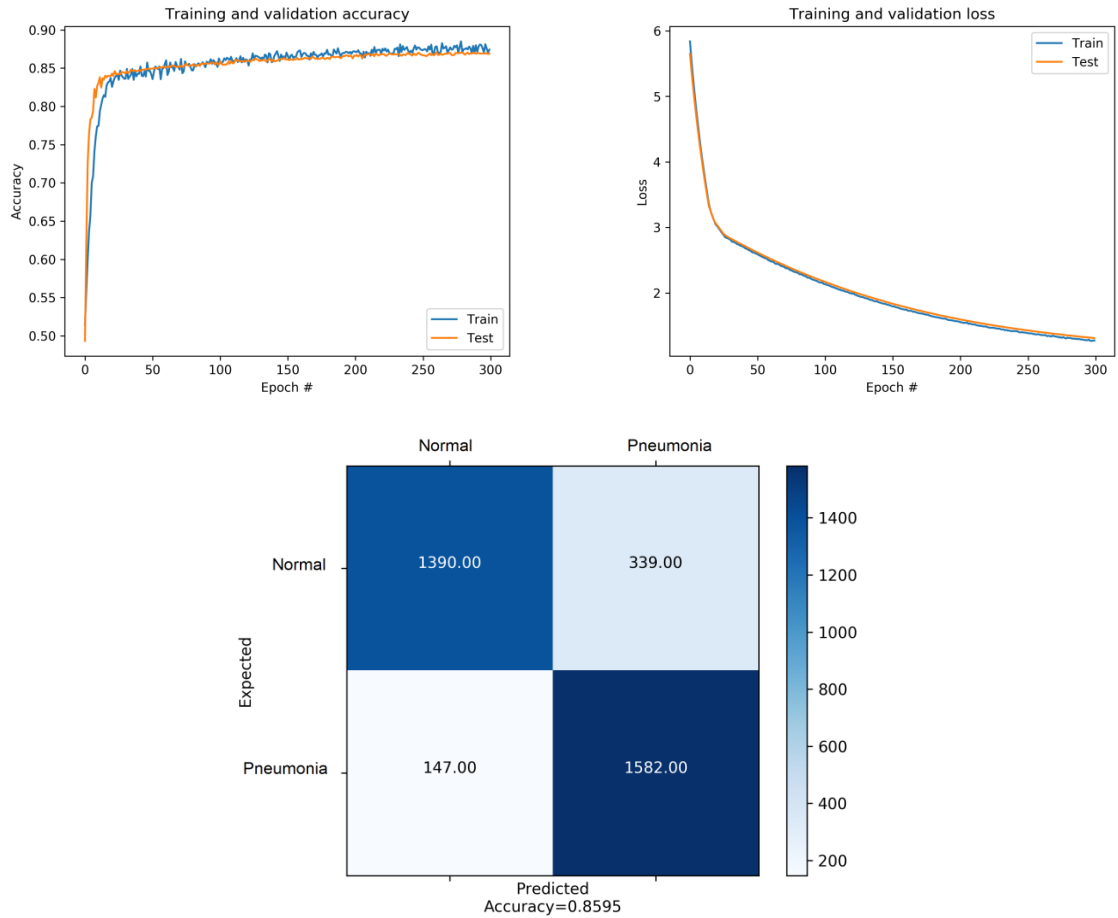

**Fig. 8**. Accuracy and loss curve and confusion matrix of VGG19

As it is showed in the figure above, the curve of train data (test data) can be divided into two intervals: the first one starts from epoch 0 to epoch 13 (from epoch 0 to epoch 10) where we can observe a quick increase of accuracy where the accuracy is equal to 81.01% (83.05%). The second interval the accuracy starts to be stable and converges to 87.42% (86.89%).

For the loss curve of train and test data, we can observe a good fit. Indeed, from epoch 0 to epoch 18 the loss is rapidly decreasing where is equal to 1.27% then it starts to be stable until the epoch 300 where is equal to 1.31%.

As observed in the normal class, the VVG19 model was able to predict 1390 images correctly and 339 images as Pneumonia. The model also was able to classify 1582 images as Pneumonia and 147 images as Normal for the Pneumonia class.

- **Inception_V3**

The next figure shows accuracy and loss curve and confusion matrix of Inception_V3. In fact, for the train and test accuracy and from epoch 0 to epoch 7, we can observe that the accuracy is increasing until the value 91.03%. After epoch 7 the accuracy starts to be stable where is equal to 97.01% and 95.94% for training data and test data respectively.

A good fit can be observed for the loss curve of train data in either the quick increasing interval from epoch 0 to epoch 32 where the loss is 3.98% or in the other interval where the decreasing is slow and converges to 1.76%.

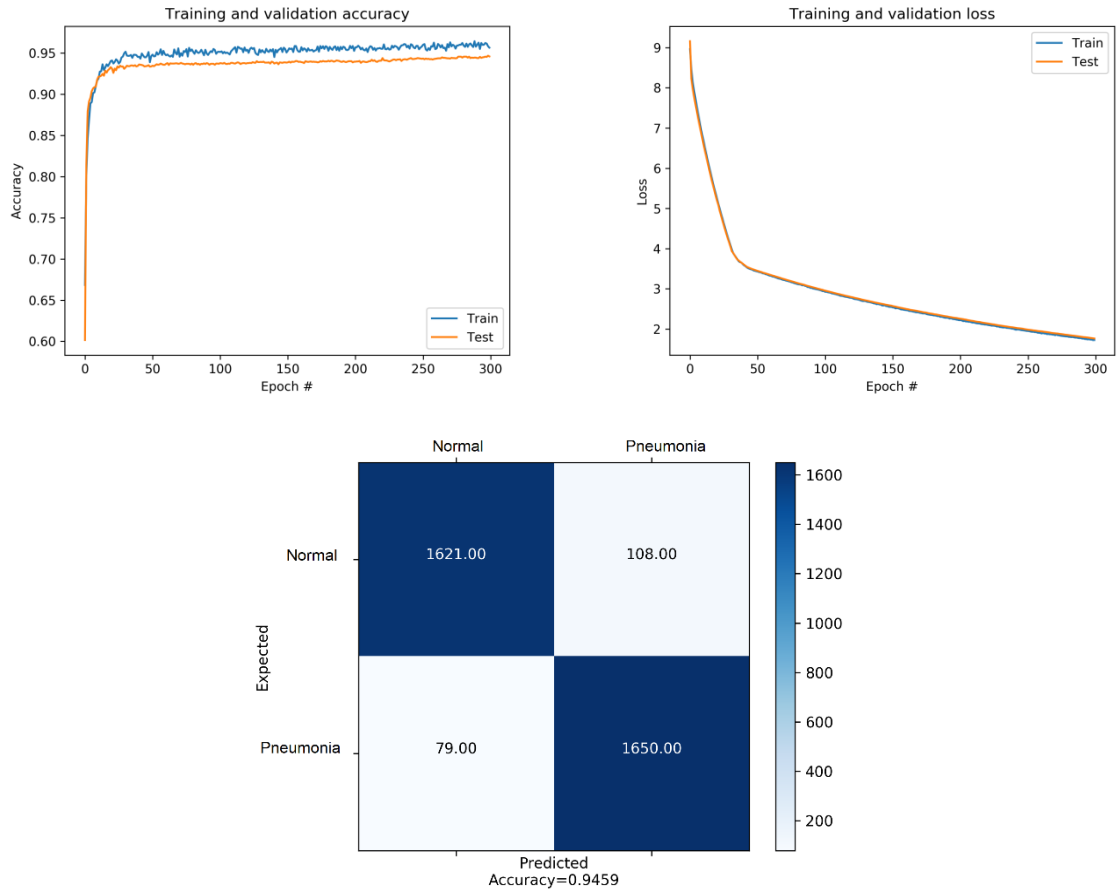

**Fig. 9**. Accuracy and loss curve and confusion matrix of Fine-tuned Inception_V3

As shown in the confusion matrix, for the Pneumonia class, the model was able to identify 1650 images as Pneumonia and 79 images as Normal. It can be seen that for images class (Normal), Inception_V3 model was able to predict 1621 images as Normal and 108 as Pneumonia.

- **ResNet50**

The figure bellow illustrates the results attained by Resnet50. In fact, from epoch 0 to epoch 24 the accuracy values are increasing expeditiously either for train data or test data where the maximum value is 97.36%. After that, the values start to be stable where the value is 99.23% and 96.23% for training data and test data respectively.

We can observe a good fit for the loss curve of train and test data, Indeed, from epoch 0 to epoch 21 the loss is briskly decreasing where is equal to 6.89% then it starts to be stable until the epoch 300 where is equal to 0.85%.

The confusion matrix indicates that, for images class (Normal), 1703 images were predicted correctly as Normal and 26 were labelled as Pneumonia. Same for the second images class (Pneumonia), the model was able to identify 1638 images correctly, but 91 images were labelled as Normal.

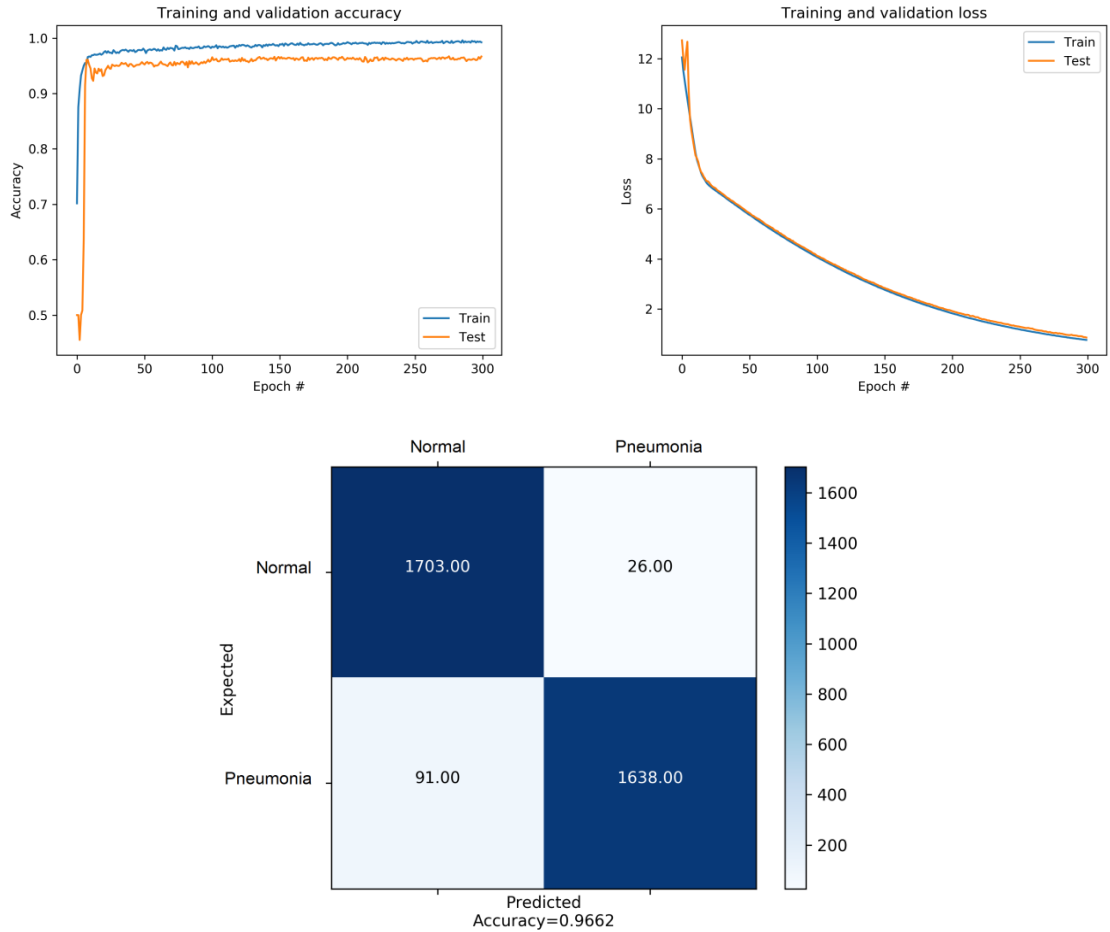

**Fig. 10**. Accuracy and loss curve and confusion matrix of Fine-tuned ResNet50

- **Inception_ResNet_V2**

From the figure bellow, we can observe that from epoch 0 to epoch 18 the train and test accuracy the accuracy is increasing until the value 95.51%. After epoch 18 the accuracy starts to be stable where is equal to 99.11% and 96.41% for training data and test data respectively.

For the loss curve, we can see an excellent fit for training and test data where the values are 3.99% (epoch 24) and 1.17% (epoch 300).

For the Pneumonia class, the Inception_ResNet_V2 model was able to identify 1618 images correctly as Pneumonia and 111 images as Normal. On other hand, for Normal class, 1705 were correctly classified as Normal and 24 images as Pneumonia.

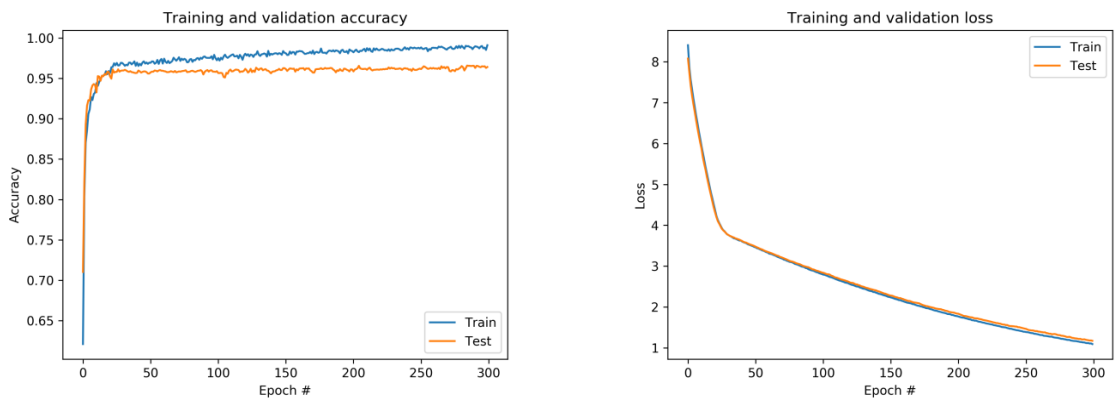

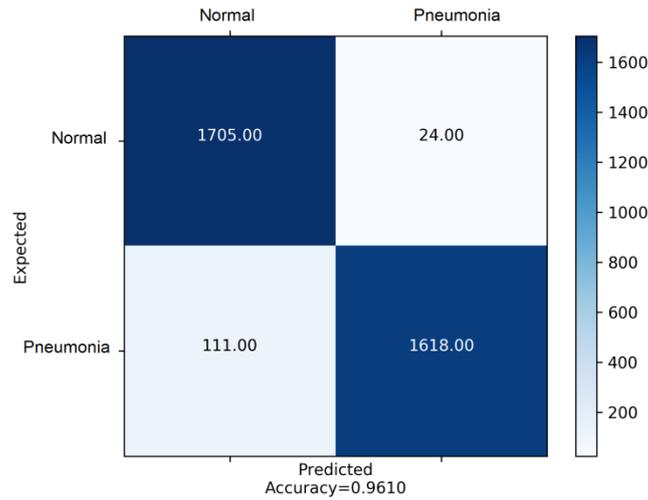

**Fig. 11**. Accuracy and loss curve and confusion matrix of Fine-tuned Inception_ResNet_V2

- **DensNet201**

The obtained accuracy curve of train data is speedily increasing until the value 93.49%. After epoch 16 the accuracy starts to be stable where is equal to 97.16% and 94.91% for train data and test data respectively.

A good fit can be checked, for the loss curve of train and test data. In fact, from epoch 0 to epoch 17 the loss is quickly decreasing where is equal to 3.99% then it starts to be stable until the epoch 300 where is equal to 1.91%.

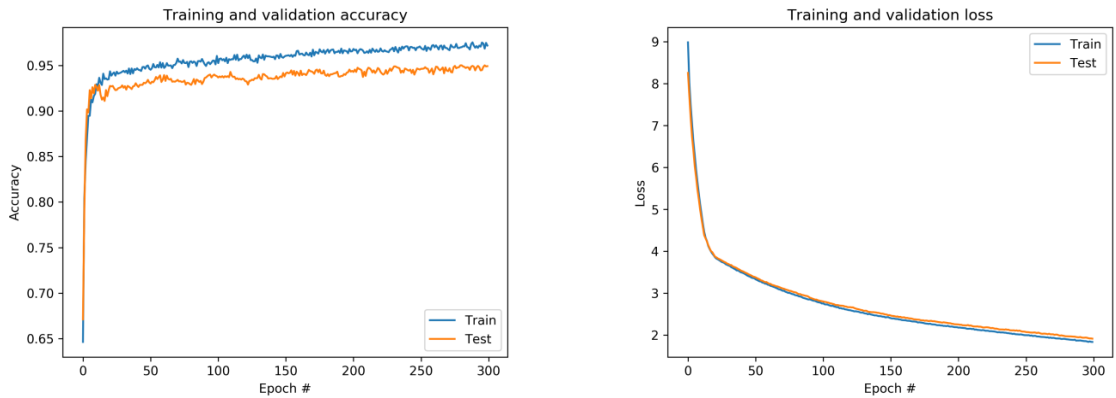

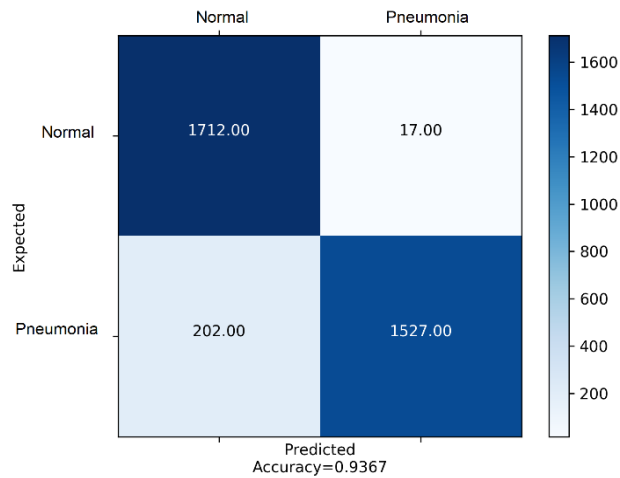

**Fig. 12**. Accuracy and loss curve and confusion matrix of Fine-tuned DensNet201

The confusion matrix depicts that for the first images class (Normal) the model was able to predict 1712 images correctly in the normal class, but 17 were labelled as Pneumonia. The model also was able to identify 1527 images correctly, but 202 images were labelled as Normal for the second images class (Pneumonia).

- **MobileNet_V2**

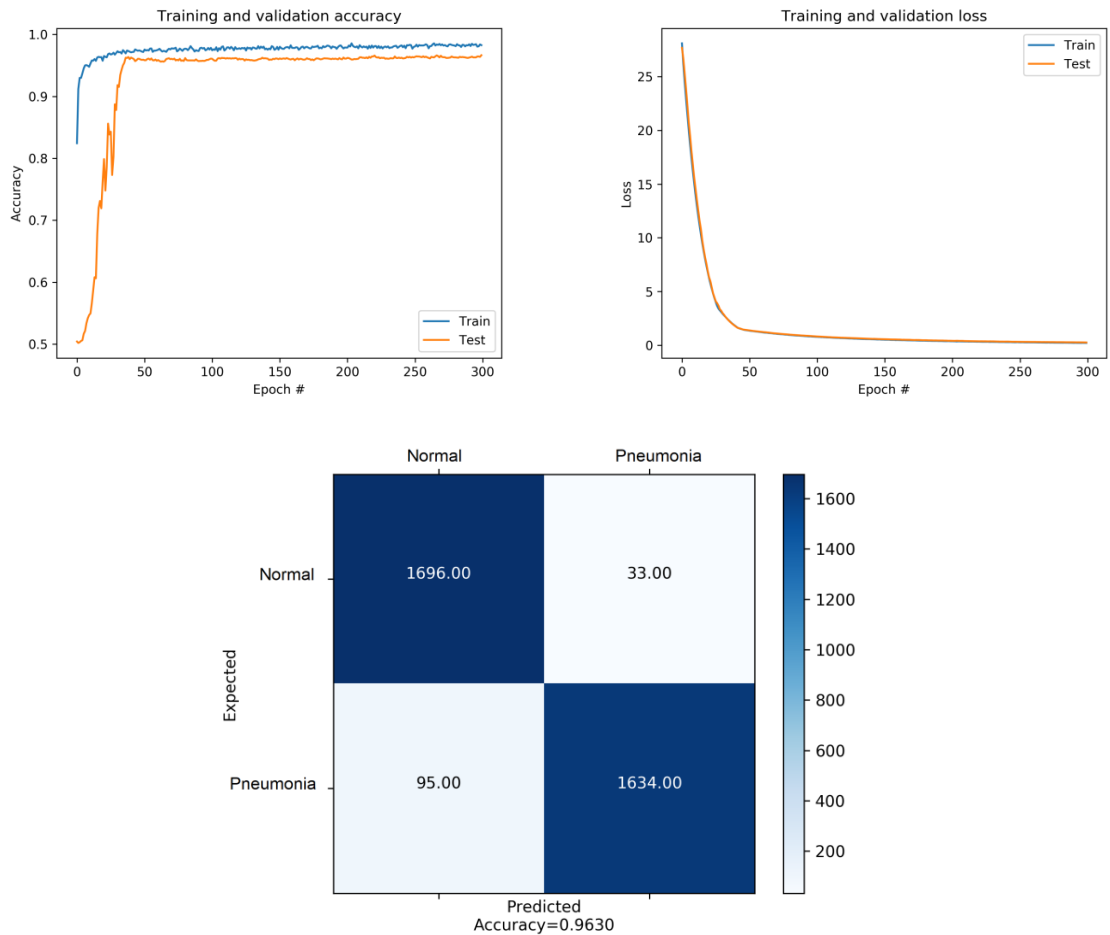

**Fig. 13**. Accuracy and loss curve and confusion matrix of Fine-tuned MobileNet_V2

As illustrated by the previous figure, we can observe that from epoch 0 to epoch 16 the train and test accuracy the accuracy is increasing until the value where the accuracy is equal to 96.38%. After epoch 16 the accuracy starts to be stable where is equal to 98.27% and 96.64% for train data and test data respectively.

For the loss curve of train data, an excellent fit is observed. Until the epoch 44 the value of loss is expeditiously decreasing where the value is 1.49% then converges to 0.24%.

Regarding to the confusion matrix above, for the first images class (Normal) the model was able to identify 1696 images correctly in the normal class, but 33 were classified as Pneumonia. Likewise, in the Pneumonia class, 1634 images were labelled correctly as Pneumonia and 95 were identified as Normal.

- **Xception**

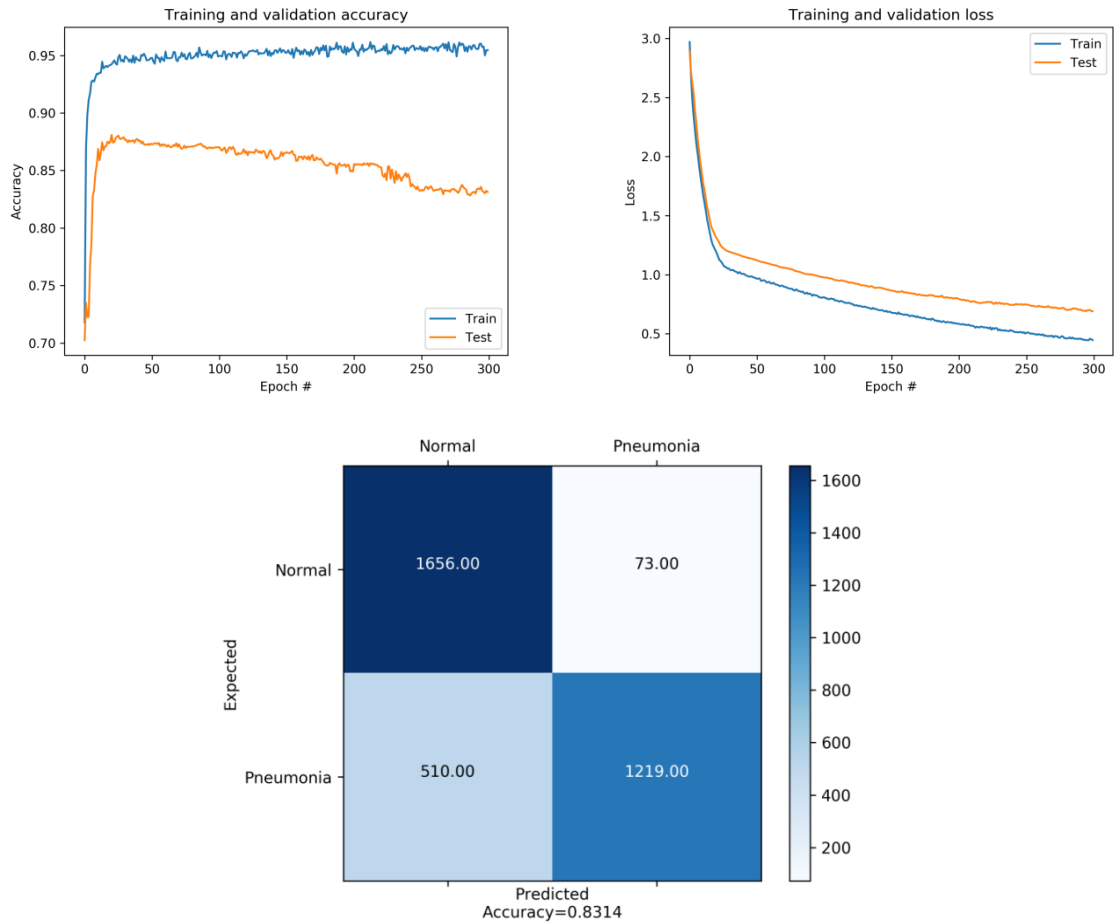

**Fig. 14**. Accuracy and loss curve and confusion matrix of Fine-tuned Xception

It is noted that the accuracy of train data is fastly increasing from epoch 0 to epoch 10 where the accuracy is equal to 93.10% then it starts to be stable until epoch 300 where the accuracy is equal to 95.45%. For the test data, a quick increasing can be observed from epoch 0 to epoch 12 where the value is 86.87% then it starts to decrease till 69.03% for epoch 300.

For the loss curve of train and test data, the values are rapidly decreasing from epoch 0 to epoch 26 where the value is 1.10%. After epoch 26 the value converges to 0.44% and 0.69% for train and test data respectively.

When we see this confusion matrix, we can observe that for the first images class (Normal), the model was able to predict 1656 images correctly in the normal class, but 73 were labelled as Pneumonia. The model also was able to identify 1219 images correctly, but 510 images were labelled as Normal for the second images class (Pneumonia).

**Discussion**

We investigated in this study the binary classification (Normal and pneumonia) based on X-Ray images using transfer learning of recent deep learning architectures, in order to identify the best performing architecture based on the several parameters defined in equation2. First, we individually compare the deep learning architectures by measuring the accuracies. After that, we compare the accuracy and loss results given by each DL architecture to discern the outperforming architecture (Fig 14, 15). Table 3 illustrates a comparison between our different deep learning models used in our experiment in terms of parameters defined in equation2.

For each model in Fig 19 (a) and (b), the plots of training and validation accuracy increase to a point of stability. It is observed that fine-tuned version of Inception_Resnet_V2, Inception_V3, Resnet50, Densnet201 and Mobilenet_V2 show highly satisfactory performance with rate of increase in training and validation accuracy with each epoch. They outperform the baseline CNN, Xception, VGG16 and VGG16 that demonstrate low performance. In fact, from epoch 20, they start to be stable until the end of training where the training and validation accuracy of baseline CNN, VGG16 and VGG16 are equals to 85%. However, Xception reaches 83 % in validation accuracy and 95% in training accuracy. In this case, the predictive model produced by Xception algorithm does not adapt well to the training Set (Overfiting). Besides, the plots of training and validation loss (Fig 20 (a) and (b)) decrease to a point of stability for each proposed model. As it can be seen, fine-tuned version of our models shows highly satisfactory performance with rate of decrease in training and validation loss with each epoch.

Results for our multi-experiment classification, based on different fine-tuned versions of recent deep learning architectures, are tabulated in Table 3. The table depicts in detail classification performances across each experiment. From the results, it's notable that the accuracy when we use baseline CNN, Xception, VGG16 and VGG19 is low compared with other DL architectures, since these last models help to obtain respectively 84.18%, 83.14%, 86.26 % and 85.94 % of accuracy. Unlike, the highest accuracies are reported by Inception_V3, DensNet201, MobileNet_V2, Inception_Resnet_V2, and Resnet50. Therefore, these models attain better classification accuracy since they achieve respectively 94.59 %, 93.66 %, 96.27 %, 96.09 % and 96.61%.

- **Training and Validation Accuracy of our models**

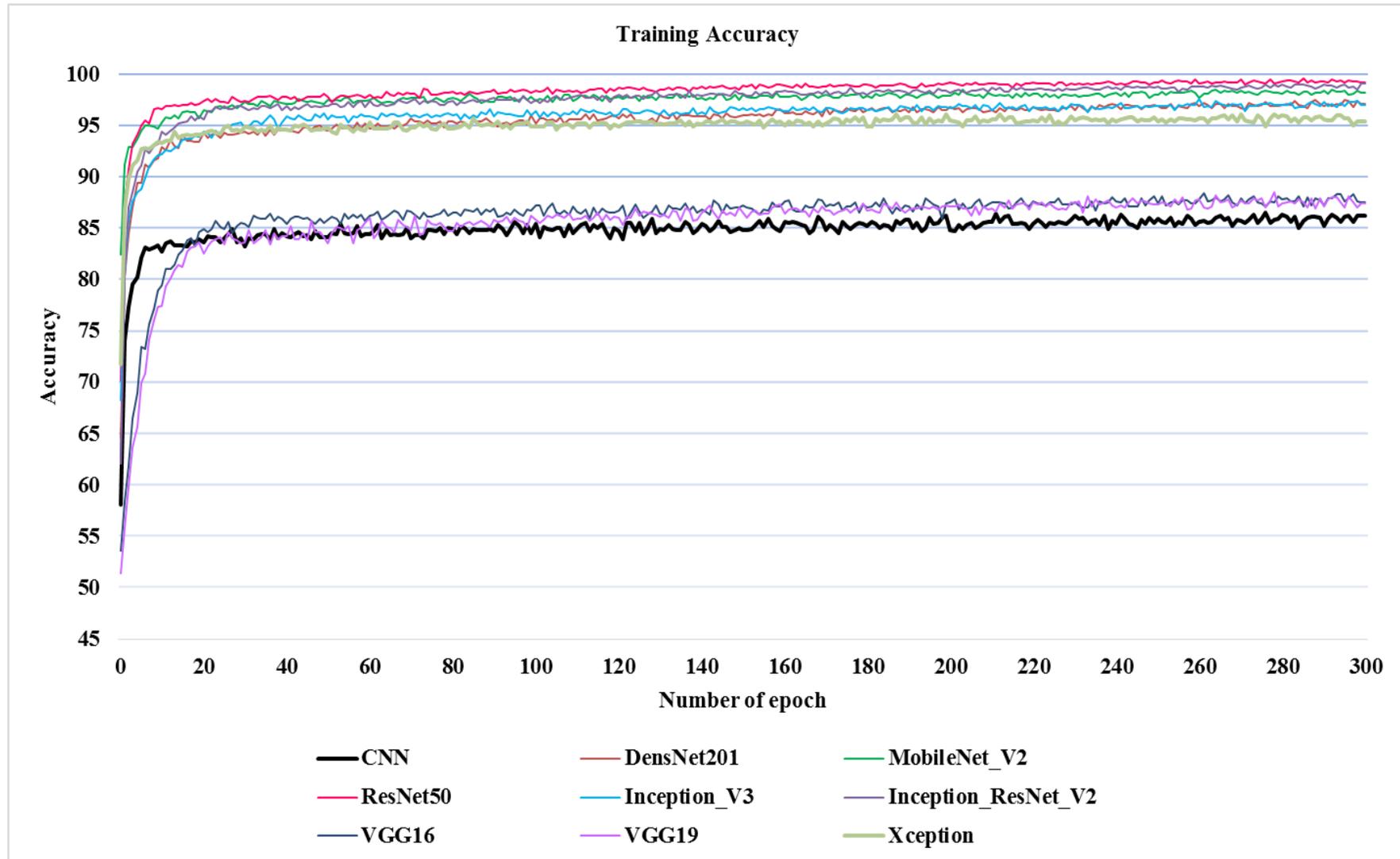

(a)

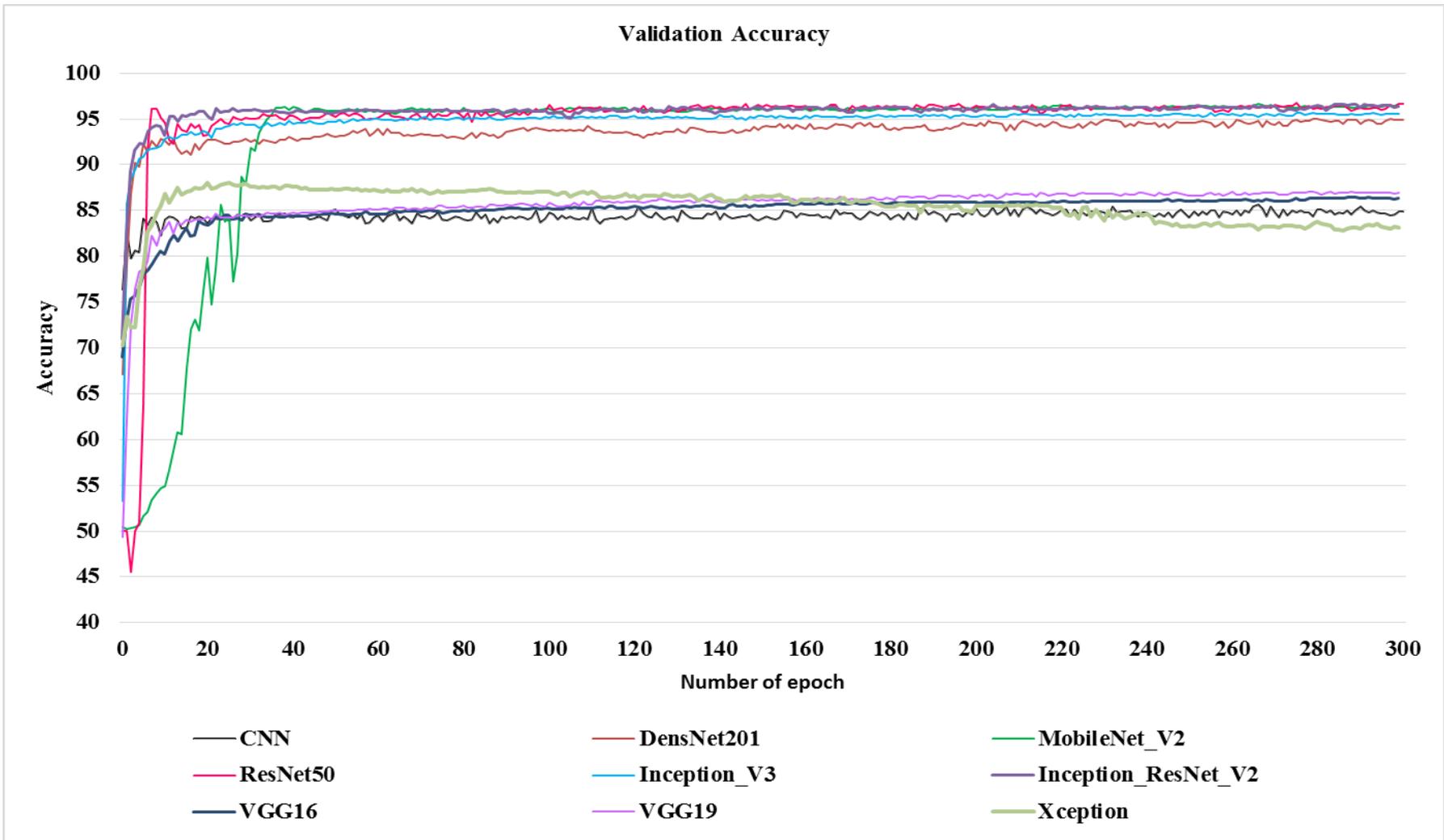

(b)

**Fig. 14**. Accuracy curve for different architectures

- **Training and Validation Loss of our models**

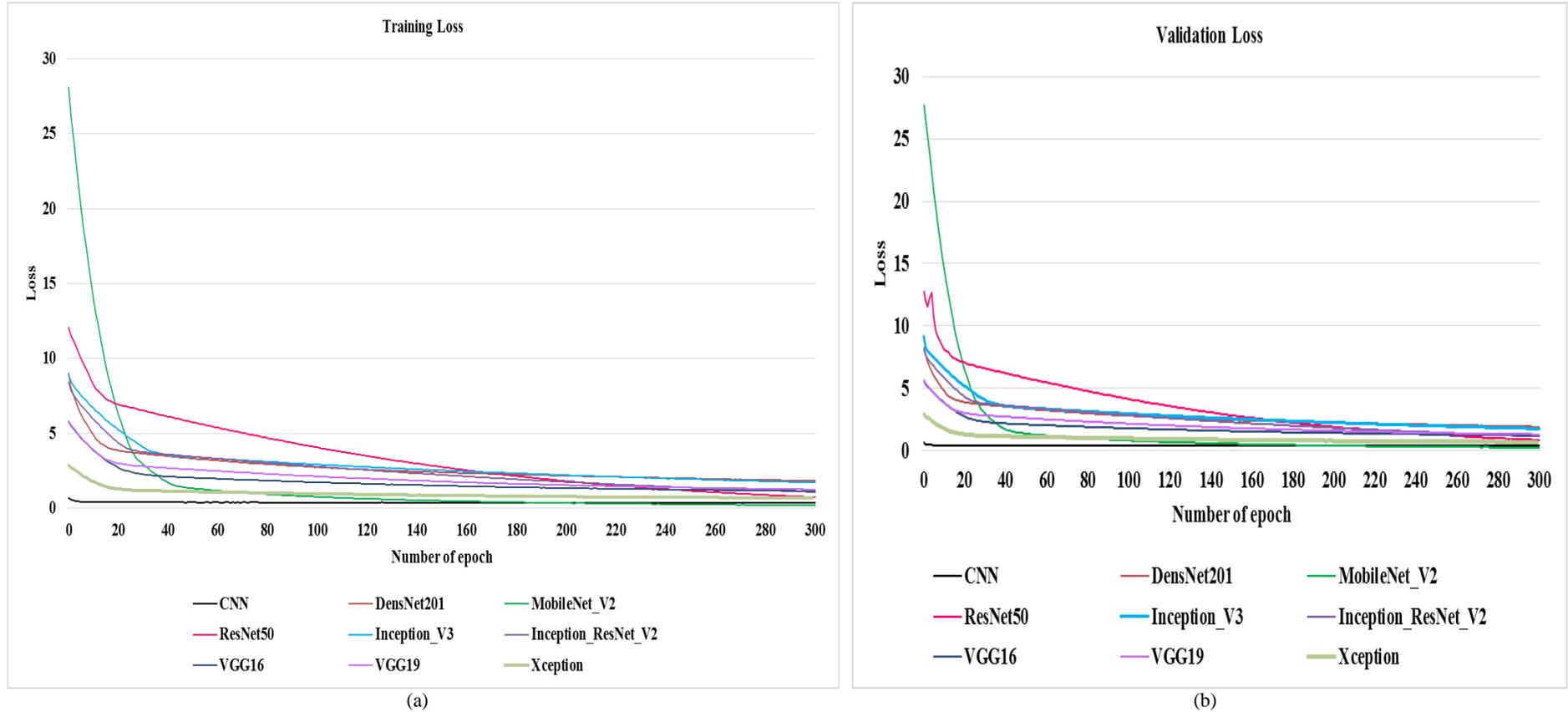

**Fig. 15**. Loss curve for different architectures

**Table 3**: Evaluation metrics

|  | TP | TN | FN | FP | Accuracy (%) | Sensitivity (%) | Specificity (%) | Precision (%) | F1 score (%) |
|---|---|---|---|---|---|---|---|---|---|
| **CNN** | 1634 | 1277 | 452 | 95 | 84.18 | 78.33 | 93.07 | 94.05 | 85.66 |
| **VGG16** | 1517 | 1466 | 263 | 212 | 86.26 | 85.22 | 87.36 | 87.73 | 86.46 |
| **VGG19** | 1390 | 1582 | 147 | 339 | 85.94 | 90.43 | 82.35 | 80.39 | 85.11 |
| **Inception_V3** | 1621 | 1650 | 79 | 108 | 94.59 | 95.35 | 93.85 | 93.75 | 94.54 |

| Model | | | | | | | | | |
|---|---|---|---|---|---|---|---|---|---|
| **Xception** | 1656 | 1219 | 510 | 73 | 83.14 | 76.45 | 94.34 | 95.77 | 8503 |
| **DensNet201** | 1712 | 1527 | 202 | 17 | 93.66 | 89.44 | 98.89 | 99.01 | 93.98 |
| **MobileNet_V2** | 1696 | 1634 | 95 | 33 | **96.27** | 94.61 | 98.02 | 98.06 | 96.30 |
| **Inception_ Resnet_V2** | 1705 | 1618 | 111 | 24 | **96.09** | 93.88 | 98.53 | 98.61 | 96.19 |
| **Resnet50** | 1703 | 1638 | 91 | 26 | **96.61** | 94.92 | 98.43 | 98.49 | 96.67 |

# V. Conclusions & Future works

We presented in this work automated methods used to classify the chest X-Ray into pneumonia and the normal class using nine Deep Learning architectures (a baseline CNN, VGG16, VGG19, DenseNet201, Inception_ResNet_V2, Inception_V3, Xception, Resnet50, and MobileNet_V2). The main goal is to answer to the following research question: Is there any Deep Learning techniques which distinctly outperforms other DCNN techniques? Toward this end, the experiments were conducted using chest X-Ray & CT dataset which contains 5856 images (4273 pneumonia and 1583 normal) and performances were evaluated using various performance metrics. Furthermore, the obtained results show that the Resnet50, MobileNet_V2 and Inception_Resnet_V2 gave highly performance (accuracy is more than 96%) against other architectures cited in this work (accuracy is around 84%).

Ongoing work intends to develop a full system for pneumonia via deep learning detection, segmentation, and classification. In addition, the performance may be improved using more datasets, more sophisticated feature extraction techniques based on deep learning such as You-Only-Look-Once (YOLO), and U-Net that was developed for biomedical image segmentation.